\documentclass{jpsj3}
\usepackage{txfonts}
\usepackage{color}

\title{Photoinduced Enhancement of Anisotropic Charge Correlations on Triangular Lattices with Trimers}

\author{Kenji Yonemitsu\thanks{E-mail: kxy@phys.chuo-u.ac.jp}}
\inst{Department of Physics, Chuo University, Bunkyo, Tokyo 112-8551, Japan} 

\abst{ To explore nontrivial photoinduced modulations of charge correlations, we theoretically study photoinduced dynamics in quarter-filled extended Hubbard models with competing intersite repulsive interactions on triangular lattices with trimers, where the end points are crystallographically equivalent. The exact diagonalization method is used and the time-dependent Schr\"odinger equation is numerically solved during and after photoexcitation. Time-averaged double occupancy and intersite density-density correlations can be interpreted as due to effective on-site and intersite repulsive interactions, respectively, relative to transfer energies. In the case where the intersite repulsive interactions compete with each other, the anisotropy of their effective interactions can be enhanced with the help of the trimers, irrespective of whether the trimers are linear or bent. In particular, in the case where the arrangement of the trimers is close to that in $\alpha$-(bis[ethylenedithio]-tetrathiafulvalene)$_2$I$_3$ [$\alpha$-(BEDT-TTF)$_2$I$_3$] in the metallic phase, the effective on-site repulsion is enhanced relative to the transfer energies. The relevance of this theoretical finding to the experimentally observed optical freezing of charge motion is discussed. }

\begin{document}
\maketitle

\section{Introduction}
Various photoinduced cooperative phenomena including photoinduced phase transitions have been studied intensively for more than a quarter century.\cite{koshigono_jpsj06,yonemitsu_pr08,basov_rmp11,nicoletti_aop16} When any order is involved, in most cases, the transition is toward destroying or melting the order. This tendency is natural because the external field supplies energy to the system and finally increases the temperature of the system. However, in transient states, the tendency can be the opposite: some order can be constructed and the electronic motion can be frozen. 

In fact, for $\alpha$-(bis[ethylenedithio]-tetrathiafulvalene)$_2$I$_3$ [$\alpha$-(BEDT-TTF)$_2$I$_3$] in a metallic phase just above the charge-order\cite{seo_jpsj00} metal-insulator transition temperature, an intense infrared pulse leads to a reflectivity change as if the electronic motion is frozen and the system transiently becomes a charge-ordered insulator.\cite{ishikawa_ncomms14} It is in contrast to the previously observed, photoinduced melting of the charge order in the same\cite{iwai_prl07,tanaka_jpsj10,miyashita_jpsj10,kawakami_prl10,gomi_jpsj11} and similar\cite{hashimoto_jpsj14,hashimoto_jpsj15} compounds. As for frozen motion, dynamical localization is theoretically known in the case of continuous waves.\cite{dunlap_prb86,grossmann_prl91,kayanuma_pra94} As a related phenomenon, a negative-temperature state is produced by continuous waves,\cite{tsuji_prl11} asymmetric pulses (like half-cycle pulses),\cite{tsuji_prb12} and symmetric pulses,\cite{yonemitsu_jpsj15,yanagiya_jpsj15} although their mechanisms are different. As for the similarity and difference between continuous-wave- and pulse-induced phenomena, the photoinduced transition from a charge-ordered-insulator phase to a Mott-insulator phase in the quasi-two-dimensional metal complex Et$_2$Me$_2$Sb[Pd(dmit)$_2$]$_2$ (dmit = 1,3-dithiol-2-thione-4,5-dithiolate)\cite{ishikawa_prb09,nishioka_jpsj13a,nishioka_jpsj13b} can be controlled in principle by suppressing interdimer electron transfer in the case of continuous waves, and by suppressing intradimer electron transfer in the case of pulses.\cite{nishioka_jpsj14} The issue on the similarity and difference between continuous-wave- and pulse-induced phenomena is also discussed in a one-dimensional transverse Ising model.\cite{ono_prb16} In this context, transient localization is expected to be caused even by a pulse, as dynamical localization is caused by a continuous wave. 

Quite recently, photoinduced changes in the direction of strengthening or enhancing the order have been observed experimentally in different situations. A tendency towards the transient stabilization of a charge density wave is demonstrated for RTe$_3$ (R=Dy, Ho) after near-infrared excitation,\cite{boven_ncomms16} counteracting the suppression of order in the nonequilibrium state, owing to transiently enhanced Fermi surface nesting. Ultrafast photoexcitation can transiently enhance the charge-density-wave amplitude in a crystalline Cr film,\cite{singer_prl16} resulting from dynamic electron-phonon interactions. The quasi-one-dimensional organic conductor (TMTTF)$_2$AsF$_6$ (TMTTF=tetramethyltetrathiafulvalene) is irradiated with 1.5-cycle, 7-fs infrared pulses, which shows an increase in the effective mass, as a strong-light field effect assisted by Coulomb repulsion.\cite{naitoh_prb16} For the layered semiconductor Ta$_2$NiSe$_5$, above a critical photoexcitation density, the direct band gap is transiently enhanced,\cite{mor_arx} owing to its exotic low-temperature ordered state. As to how intense THz excitations can be used to resonantly control matter, through superconducting gaps, Josephson plasmons in layered superconductors, vibrational modes of the crystal lattice, and magnetic excitations, see Ref.~\citen{nicoletti_aop16}. 

In this paper, bearing in mind the optical freezing of charge motion in $\alpha$-(BEDT-TTF)$_2$I$_3$ in the metallic phase, we theoretically discuss the feasibility of pulse-induced transient localization and modulations of charge correlations by ultrafast and intense photoexcitation. Its relationship with the lattice structure (i.e., the network of transfer integrals) and interactions will be clarified. For $\alpha$-(BEDT-TTF)$_2$I$_3$, competing intersite repulsive interactions and the lattice with trimers possessing crystallographically equivalent end points are shown to be the key ingredients for the observed phenomena. Furthermore, it is shown that the photoinduced enhancement of anisotropic charge correlations is not limited to a lattice structure similar to that of $\alpha$-(BEDT-TTF)$_2$I$_3$: it is observed regardless of whether the trimers are linear or bent. 

\section{Extended Hubbard Models on Triangular Lattices with Trimers \label{sec:model}}
\begin{figure}
\includegraphics[height=15.2cm]{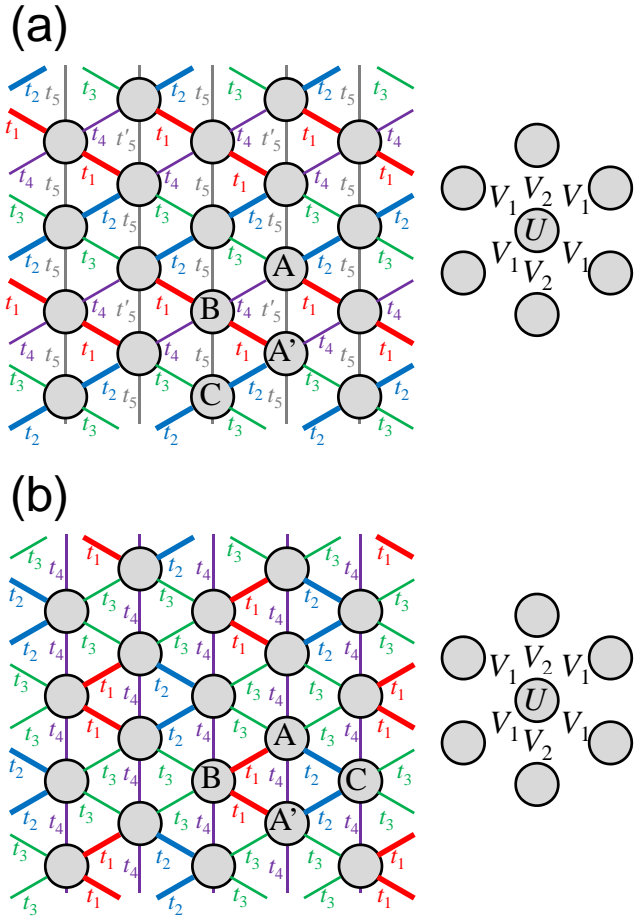}
\caption{(Color online) 
Triangular lattices with (a) linear and (b) bent trimers. The lattice (a) has inversion symmetry, while (b) has reflection symmetry. Sites A and A' are crystallographically equivalent.
\label{fig:latt_str}}
\end{figure}

We use quarter-filled extended Hubbard models, 
\begin{equation}
H = 
\sum_{\langle ij \rangle \sigma}
t_{ij}  ( c^\dagger_{i\sigma} c_{j\sigma} + c^\dagger_{j\sigma} c_{i\sigma}  )
+U\sum_i n_{i\uparrow} n_{i\downarrow}
+\sum_{\langle ij \rangle} V_{ij} n_i n_j
\;, \label{eq:model}
\end{equation}
where $ c^\dagger_{i\sigma} $ creates an electron with spin $ \sigma $ at site $ i $, $ n_{i\sigma} $=$ c^\dagger_{i\sigma} c_{i\sigma} $, and $ n_i $=$ \sum_\sigma n_{i\sigma} $. The parameter $ U $ represents the on-site Coulomb repulsion. We consider the triangular lattice that consists of equilateral triangles, where the distance between neighboring sites is denoted by $a$. The location of the $ i $th site is represented by $ \mbox{\boldmath $r$}_i $, and the relative position by $ \mbox{\boldmath $r$}_{ij}=\mbox{\boldmath $r$}_j-\mbox{\boldmath $r$}_i $. The nearest-neighbor Coulomb repulsion $ V_{ij} $ is assumed to be $V_{ij}=V_1$ for $ \mbox{\boldmath $r$}_{ij}$ being at a $\pm$30 or $\pm$150 -degree angle with respect to the horizontal axis, and $V_{ij}=V_2$ for $ \mbox{\boldmath $r$}_{ij}$ being parallel to the vertical axis, as shown in Fig.~\ref{fig:latt_str}. 
For the interaction parameters, we use $U=0.8$ and $V_1=0.3$, and vary $V_2$. 

For the transfer integrals $t_{ij}$, we consider the two cases shown in Figs.~\ref{fig:latt_str}(a) and \ref{fig:latt_str}(b). The structure in Fig.~\ref{fig:latt_str}(a) has linear trimers and is close to that of $\alpha$-(BEDT-TTF)$_2$I$_3$ in the metallic phase, although the latter does not consist of equilateral triangles. The structure in Fig.~\ref{fig:latt_str}(b) has bent trimers. The three sites (A, B, and A' in Fig.~\ref{fig:latt_str}) in a trimer are linked by the largest transfer integral in magnitude, $t_1$. There is another type of trimer, ``weak trimer,'' where the three sites (A, C, and A' in Fig.~\ref{fig:latt_str}) are linked by the second largest transfer integral in magnitude, $t_2$. The end points of a weak trimer (A and A') are end points of one [Fig.~\ref{fig:latt_str}(b)] or two [Fig.~\ref{fig:latt_str}(a)] ``strong trimers.'' The structure in Fig.~\ref{fig:latt_str}(a) has inversion symmetry with respect to site B, site C, or the midpoint of neighboring sites A and A'. This symmetry is possessed by $\alpha$-(BEDT-TTF)$_2$I$_3$ in the metallic phase. The structure in Fig.~\ref{fig:latt_str}(b) has reflection symmetry with respect to a line parallel to the horizontal axis. The notations of sites A, A', B, and C are the same as those in Refs.~\citen{tanaka_jpsj10} and \citen{miyashita_jpsj10} for $\alpha$-(BEDT-TTF)$_2$I$_3$. In this paper, we use $t_1=-0.14$, $t_2=-0.13$, $t_3=-0.02$, $t_4=-0.06$, $t_5=0.03$, and $t_5'=-0.03$ for Fig.~\ref{fig:latt_str}(a), and $t_1=-0.10$, $t_2=-0.08$, $t_3=-0.02$, and $t_4=-0.01$ for Fig.~\ref{fig:latt_str}(b). The parameters for Fig.~\ref{fig:latt_str}(a) are close to those of $\alpha$-(BEDT-TTF)$_2$I$_3$ in the metallic phase. However, the conclusions of this study do not depend on the details in these parameters. Note in Fig.~\ref{fig:latt_str}(a) that sites A and A' along the vertical axis are alternately linked by $t_5$ and $t_5'$, as in $\alpha$-(BEDT-TTF)$_2$I$_3$, which leads to similar $V_1$ and $V_2$ dependences of the electron density distributions for the networks of transfer integrals in Figs.~\ref{fig:latt_str}(a) and \ref{fig:latt_str}(b) and in $\alpha$-(BEDT-TTF)$_2$I$_3$ in the metallic phase. Note also that $\alpha$-(BEDT-TTF)$_2$I$_3$ is a three-quarter-filled system, i.e., a quarter-filled system in the hole picture. The transfer integrals in the hole picture are inverted from those in the electron picture here. 

The initial state is the ground state obtained by the exact diagonalization method for the 16-site system with periodic boundary conditions. Photoexcitation is introduced through the Peierls phase 
\begin{equation}
c_{i\sigma}^\dagger c_{j\sigma} \rightarrow
\exp \left[
\frac{ie}{\hbar c} \mbox{\boldmath $r$}_{ij} \cdot \mbox{\boldmath $A$}(t)
\right] c_{i\sigma}^\dagger c_{j\sigma}
\;, \label{eq:photo_excitation}
\end{equation}
which is substituted into Eq.~(\ref{eq:model}). We employ symmetric monocycle electric-field pulses\cite{yonemitsu_jpsj15,yanagiya_jpsj15} and use the time-dependent vector potential 
\begin{equation}
\mbox{\boldmath $A$} (t) = \frac{c\mbox{\boldmath $F$}}{\omega} \left[ \cos (\omega t)-1 \right] 
\theta (t) \theta \left( \frac{2\pi}{\omega}-t \right)
\;, \label{eq:monocycle_pulse}
\end{equation}
with $ \mbox{\boldmath $F$}=F(\cos \theta,\sin \theta) $, where $F$ is the amplitude of the electric field and $\theta$ is the angle between the field and the horizontal axis. The central frequency $ \omega $ is chosen to be $ \omega=0.8 $, which is well above the main charge-transfer excitations, as shown below. The optical conductivity spectra are calculated as before.\cite{miyashita_jpsj10} The time-dependent Schr\"odinger equation is numerically solved by expanding the exponential evolution operator with a time slice $ dt $=0.02 to the 15th order and by checking the conservation of the norm.\cite{yonemitsu_prb09} The time average $\langle \langle Q \rangle \rangle $ of a quantity $Q$ is calculated as 
\begin{equation}
\langle \langle Q \rangle \rangle =
\frac{1}{t_w} \int_{t_s}^{t_s+t_w} 
\langle \Psi (t) \mid Q \mid \Psi (t) \rangle dt
\;, \label{eq:time_average}
\end{equation}
with $ t_s=5 T $, $ t_w=5 T $, and $ T=2\pi/\omega $ being the period of the oscillating electric field. If the model parameters are represented in eV, $T$ corresponds to about 5 fs, and $ t_s+t_w = 10T $ corresponds to about 52 fs. Qualitative results are unchanged even if different time intervals are used. 

\section{Triangular Lattice with Linear Trimers}

\subsection{Ground states with inversion symmetry}
\begin{figure}
\includegraphics[height=16.8cm]{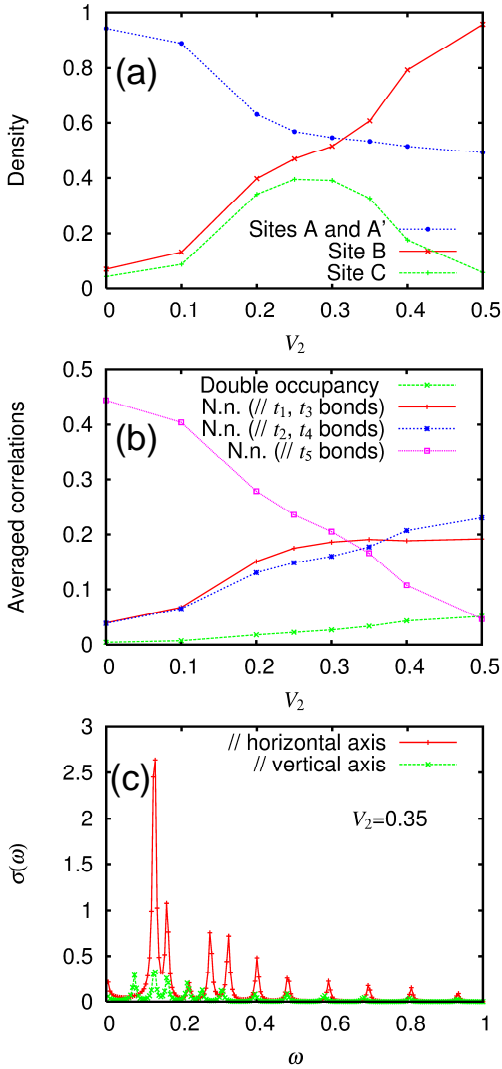}
\caption{(Color online) 
(a) Electron density $\langle n_i \rangle$ at sites A, A', B, and C, and 
(b) spatially averaged, double occupancy $\langle n_{i\uparrow} n_{i\downarrow} \rangle$, nearest-neighbor density-density correlation $\langle n_i n_j \rangle$ for $ \mbox{\boldmath $r$}_{ij}$ at a $-$30 -degree angle with respect to horizontal axis (i.e., parallel to $t_1$ and $t_3$ bonds), $\langle n_i n_j \rangle$ for $ \mbox{\boldmath $r$}_{ij}$ at a 30 -degree angle (i.e., parallel to $t_2$ and $t_4$ bonds), and  $\langle n_i n_j \rangle$ for $ \mbox{\boldmath $r$}_{ij}$ parallel to vertical axis (i.e., parallel to $t_5$ and $t_5'$ bonds), as functions of $V_2$. (c) Optical conductivity spectra for $V_2=0.35$ with different polarizations as indicated. The peak-broadening parameter is set at 0.005. 
\label{fig:lnr_atr16_trat1413020603u8v3vxx_eq}}
\end{figure}

Here, we consider the triangular lattice with linear trimers shown in Fig.~\ref{fig:latt_str}(a). To see how the intersite repulsive interactions $V_1$ and $V_2$ govern the ground states, we show the electron density distribution $\langle n_i \rangle$ in Fig.~\ref{fig:lnr_atr16_trat1413020603u8v3vxx_eq}(a) and the spatially averaged correlation functions $\langle n_{i\uparrow} n_{i\downarrow} \rangle$ and $\langle n_i n_j \rangle$, which contribute to the interaction energies, in Fig.~\ref{fig:lnr_atr16_trat1413020603u8v3vxx_eq}(b), as functions of $V_2$. 
Note that the values and symmetry of the transfer integrals used here are close to those of $\alpha$-(BEDT-TTF)$_2$I$_3$ in the metallic phase. In the present finite-size systems, the symmetry is not spontaneously broken, so that the ground states have the same symmetry as $\alpha$-(BEDT-TTF)$_2$I$_3$ has in the metallic phase. Sites A and A' are crystallographically equivalent, so that their electron densities are equal in the ground states. The on-site repulsion $U$ is the largest parameter, so that the double occupancy $\langle n_{i\uparrow} n_{i\downarrow} \rangle$ is small over the entire region shown in this figure. 

When $V_2$ is small, $V_1$ dominates the ground state, and the nearest-neighbor density-density correlations $\langle n_i n_j \rangle$ are small along the $t_1$, $t_2$, $t_3$, and $t_4$ bonds. In this case, the electron density should be distributed mainly to, in principle, either sites A and A' or sites B and C along the $t_5$ and $t_5'$ bonds. For the present choice of the transfer integrals, electrons are mainly on sites A and A'. As $V_2$ increases, electrons on neighboring sites A and A' repel each other more strongly, but the symmetry is unchanged, so that their electron densities remain equal. They simply decrease and the electron densities increase at sites B and C. 

When $V_2$ becomes comparable to $V_1$(=0.3), the spatial variation of $\langle n_{i\uparrow} n_{i\downarrow} \rangle$ and that of $\langle n_i n_j \rangle$ become small. In this case, their distributions are governed by the transfer integrals, i.e., they are determined in such a way that the kinetic energy is lowered:  the electron densities at sites A, B, and A' on trimers are slightly larger than that at site C, and $\langle n_i n_j \rangle$ along the $t_1$ bond is slightly larger than $\langle n_i n_j \rangle$ along the $t_2$ bond. This situation is realized in $\alpha$-(BEDT-TTF)$_2$I$_3$ in the metallic phase: the charge disproportionation at sites B and C is due to this network of transfer integrals. As $V_2$ further increases, the electron densities at sites A, B, and A' on a trimer approach 0.5, 1, and 0.5, respectively, maintaining the symmetry with respect to the exchange of sites A and A'. 

Hereafter, we mainly use $V_2=0.35$ for photoinduced dynamics. Then, we show the optical conductivity spectra for $V_2=0.35$ in Fig.~\ref{fig:lnr_atr16_trat1413020603u8v3vxx_eq}(c). The main charge-transfer excitations are seen below 0.4, so that $ \omega=0.8 $ for photoexcitation used below is well above them. 

\subsection{Densities and correlations after pulse excitation \label{sec:linear}}
Densities and correlation functions that are time-averaged after the monocycle pulse excitation as explained in Sect.~\ref{sec:model} are shown below, as functions of the dimensionless quantity equal to the ratio of the electric field amplitude to the central frequency $ eaF/(\hbar\omega) $. Unless stated otherwise, we use $V_2=0.35$ and $\theta=0$ in Sect.~\ref{sec:linear}. 

\begin{figure}
\includegraphics[height=16.8cm]{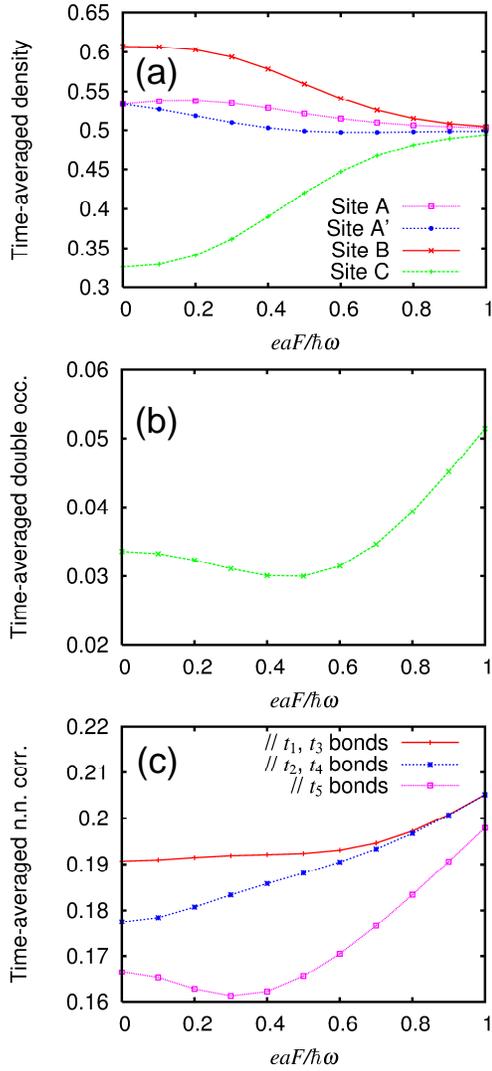}
\caption{(Color online) 
(a) Time-averaged electron densities $\langle \langle n_i \rangle \rangle$ at sites A, A', B, and C, (b) spatially and temporally averaged double occupancy $\langle \langle n_{i\uparrow} n_{i\downarrow} \rangle \rangle$, and (c) spatially and temporally averaged nearest-neighbor density-density correlations $\langle \langle n_i n_j \rangle \rangle$ for $ \mbox{\boldmath $r$}_{ij}$ in different directions as indicated, for $V_2=0.35$ and $\theta=0$, as functions of $ eaF/(\hbar\omega) $. 
\label{fig:lnr_atr16_trat1413020603u8v3v35_w0p80n1fxq00}}
\end{figure}

The time-averaged electron densities $\langle \langle n_i \rangle \rangle$ are shown in Fig.~\ref{fig:lnr_atr16_trat1413020603u8v3v35_w0p80n1fxq00}(a). 
In the ground state before photoexcitation, electrons are mainly on trimers (i.e., sites A-B-A'). After photoexcitation, as the field amplitude increases, the charge disproportionation becomes weak, and the time-averaged electron density distribution approaches the homogeneous one, as expected. Here, the equivalence of sites A and A' is broken by the field. When the field is inverted ($\theta \rightarrow \theta+180\deg$), $\langle \langle n_i \rangle \rangle$ at site A and $\langle \langle n_i \rangle \rangle$ at site A' are exchanged. Because the charge disproportionation at sites B and C is of kinetic origin, as mentioned above, and because the 0.5-0.5 density distribution at sites A-A' can be a linear combination of the 1-0 and 0-1 density distributions, we need to see the correlation functions. 

The spatially and temporally averaged double occupancy $\langle \langle n_{i\uparrow} n_{i\downarrow} \rangle \rangle$ once decreases and then increases as the field amplitude increases, as shown in Fig.~\ref{fig:lnr_atr16_trat1413020603u8v3v35_w0p80n1fxq00}(b). The decreased $\langle \langle n_{i\uparrow} n_{i\downarrow} \rangle \rangle$ means that electrons with opposite spins avoid being on the same site more strongly as if the on-site repulsion $U$ were increased relative to the transfer integrals transiently after photoexcitation. The increased $\langle \langle n_{i\uparrow} n_{i\downarrow} \rangle \rangle$ for larger field amplitudes is natural since the total energy increases and the interaction energy due to the double occupancy contributes to the total energy. 

The spatially and temporally averaged nearest-neighbor density-density correlations $\langle \langle n_i n_j \rangle \rangle$ show anisotropic behavior, as shown in Fig.~\ref{fig:lnr_atr16_trat1413020603u8v3v35_w0p80n1fxq00}(c). Because of $V_2 > V_1$, $\langle \langle n_i n_j \rangle \rangle$ along the $t_5$ and $t_5'$ bonds is smaller than the others already before photoexcitation (at $F=0$), and it is further decreased by not-too-strong fields [$ eaF/(\hbar\omega) \leq 0.5 $]. This behavior means that electrons avoid neighboring along the $t_5$ and $t_5'$ bonds more strongly as if the intersite repulsive interaction $V_2$ were increased relative to the transfer integrals transiently after photoexcitation. Meanwhile, $\langle \langle n_i n_j \rangle \rangle$ along the $t_1$ and $t_3$ bonds and $\langle \langle n_i n_j \rangle \rangle$ along the $t_2$ and $t_4$ bonds increase with the field amplitude. Thus, the anisotropy in $\langle \langle n_i n_j \rangle \rangle$ is enhanced by electric-field pulses, which is interpreted as the photoinduced enhancement of the anisotropy in the {\it effective} intersite repulsive interactions. 

The present ground state corresponds to $\alpha$-(BEDT-TTF)$_2$I$_3$ in the metallic phase once electrons in the former are regarded as holes in the latter. Note that, in equilibrium, the charge order in the latter below the transition temperature is mainly caused by the intersite repulsive interaction $V_2$ ($> V_1$) and is assisted by the coupling to the lattice degrees of freedom.\cite{tanaka_jpsj08,miyashita_jpsj08} An infrared pulse causes a reflectivity change as if the system becomes a charge-ordered insulator for 50 fs.\cite{ishikawa_ncomms14} The short lifetime is possibly due to a short-range nature of the order. This experimental fact is consistent with the present theoretical result in the sense that the enhanced {\it effective} interactions $U$ and $V_2$ relative to the transfer integrals can support a short-range charge order since the observed state is close to the metal-insulator phase boundary. The numerical calculations are performed for exact many-electron wave functions on finite-size systems that are not coupled to lattice degrees of freedom, so that the symmetry cannot be broken spontaneously but can be broken by the external field. If we could approach the thermodynamic limit, the modulations of correlation functions would be larger than the present ones. 

\begin{figure}
\includegraphics[height=5.6cm]{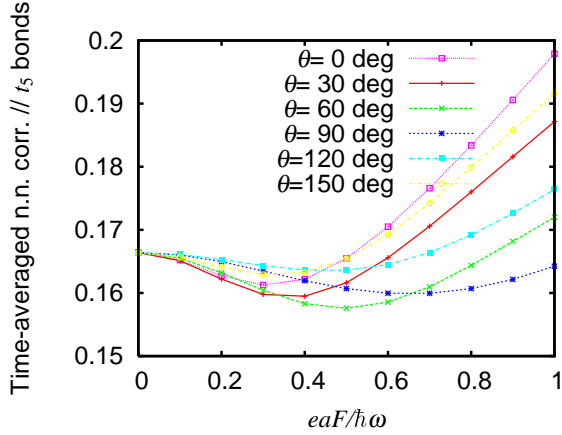}
\caption{(Color online) 
Spatially and temporally averaged nearest-neighbor density-density correlation $\langle \langle n_i n_j \rangle \rangle$ for $ \mbox{\boldmath $r$}_{ij}$ parallel to vertical axis (i.e., parallel to $t_5$ and $t_5'$ bonds), for $V_2=0.35$ and different polarizations of photoexcitation $\theta$, as functions of $ eaF/(\hbar\omega) $. 
\label{fig:lnr_atr16_trat1413020603u8v3v35_w0p80n1fxqyy_avnn}}
\end{figure}

The dependence on the polarization of photoexcitation $\theta$ is shown in Fig.~\ref{fig:lnr_atr16_trat1413020603u8v3v35_w0p80n1fxqyy_avnn} for $\langle \langle n_i n_j \rangle \rangle$ along the $t_5$ and $t_5'$ bonds. 
Note that $\theta=0$ corresponds to the polarization along the $b$-axis in $\alpha$-(BEDT-TTF)$_2$I$_3$. For small field amplitudes, $ eaF/(\hbar\omega)$=0.2 and 0.3, $\langle \langle n_i n_j \rangle \rangle$ becomes smallest at approximately $\theta=30\deg$. For larger field amplitudes, $ eaF/(\hbar\omega)$=0.4, 0.5, and 0.6,  $\langle \langle n_i n_j \rangle \rangle$ becomes smallest at approximately $\theta=60\deg$. The polarization $\theta=30\deg$ is along the $t2$ bonds in Fig.~\ref{fig:latt_str}(a), where the transfer integral is the second largest in magnitude. The corresponding electron transfer process efficiently uses site C where the electron density is lowest in the ground state. Sites A (A') and B linked by $t_1$ are already almost singly occupied and the electron transfer through this bond ($\theta=150\deg$) is hindered by the strong on-site repulsion $U$, so that the $t_1$ process is less efficient. The $t_3$ process ($\theta=150\deg$) can efficiently use site C in principle, but $t_3$ is much smaller in magnitude. Then, the $t_2$ process efficiently makes $\langle \langle n_i \rangle \rangle$ at sites A and A' approach 0.5, enhancing the 1-0/0-1 density correlation at sites A-A' on the $t_5$ and $t_5'$ bonds. The precise polarization dependence is not so simple because the network of transfer integrals is rather complex. 

\begin{figure}
\includegraphics[height=11.2cm]{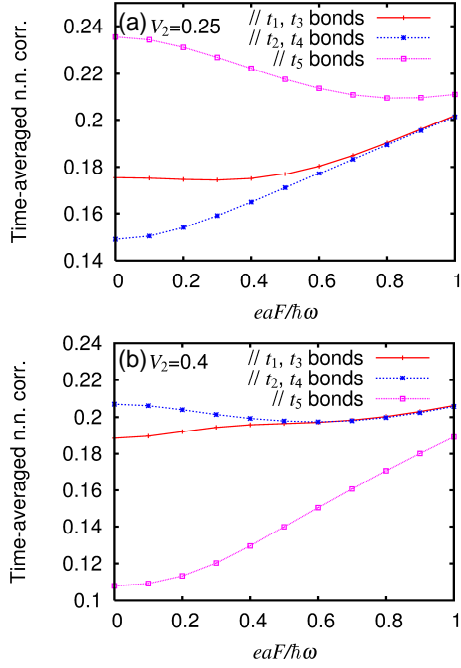}
\caption{(Color online) 
Spatially and temporally averaged nearest-neighbor density-density correlations $\langle \langle n_i n_j \rangle \rangle$ for $ \mbox{\boldmath $r$}_{ij}$ in different directions as indicated, for $\theta=0$, (a) $V_2=0.25$ and (b) $V_2=0.4$, as functions of $ eaF/(\hbar\omega) $. 
\label{fig:lnr_atr16_trat1413020603u8v3v25v4_w0p80n1fxq00_avnn3}}
\end{figure}

To see how general/special the photoinduced enhancement of the anisotropy in $\langle \langle n_i n_j \rangle \rangle$ is, $\langle \langle n_i n_j \rangle \rangle$ for $V_2=0.25$ and $V_2=0.4$ are shown in Figs.~\ref{fig:lnr_atr16_trat1413020603u8v3v25v4_w0p80n1fxq00_avnn3}(a) and \ref{fig:lnr_atr16_trat1413020603u8v3v25v4_w0p80n1fxq00_avnn3}(b), respectively. 
For $V_2=0.25$, $\langle \langle n_i n_j \rangle \rangle$ along the $t_1$, $t_2$, $t_3$, and $t_4$ bonds, which are suppressed by $V_1$ ($> V_2$) before photoexcitation, increase with the field amplitude, while $\langle \langle n_i n_j \rangle \rangle$ along the $t_5$ and $t_5'$ bonds decreases. The anisotropy in $\langle \langle n_i n_j \rangle \rangle$ is weakened by photoexcitation. For $V_2=0.4$, the roles of the interactions $V_1$ and $V_2$ are only exchanged, so that the anisotropy in $\langle \langle n_i n_j \rangle \rangle$ is also weakened here. These behaviors are in contrast to the photoinduced enhancement of the anisotropy in  $\langle \langle n_i n_j \rangle \rangle$ for $V_2=0.35$ shown in Fig.~\ref{fig:lnr_atr16_trat1413020603u8v3v35_w0p80n1fxq00}(c). The latter behavior is observed only when $V_2$ is slightly larger than $V_1$ and $\langle \langle n_i n_j \rangle \rangle$ in different directions compete with each other. The difference between the cases of $V_2 > V_1$ and $V_1 > V_2$ is due to the presence of trimers since the difference is already seen in the ground states: Fig.~\ref{fig:lnr_atr16_trat1413020603u8v3vxx_eq}(b) is not symmetric with respect to the $V_2 = V_1$ point, owing to the structure with trimers. Then, the photoinduced enhancement of anisotropic charge correlations is presumed to be caused by the competing intersite repulsive interactions on a lattice with trimers. The next issue to be resolved is how generally this occurs on different lattices with trimers. 

\section{Triangular Lattice with Bent Trimers}

\subsection{Ground states with reflection symmetry}
\begin{figure}
\includegraphics[height=16.8cm]{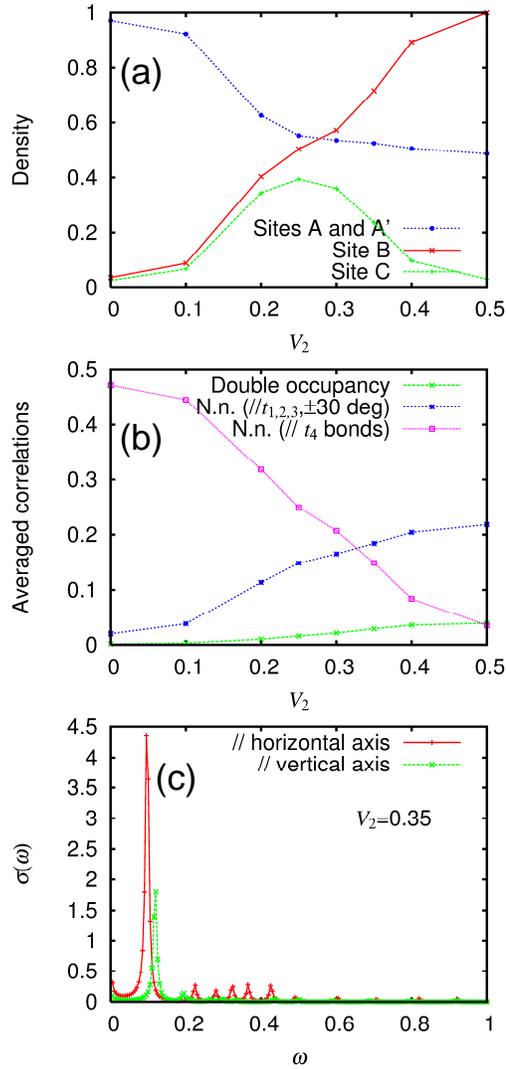}
\caption{(Color online) 
(a) Electron density $\langle n_i \rangle$ at sites A, A', B, and C, and 
(b) spatially averaged, double occupancy $\langle n_{i\uparrow} n_{i\downarrow} \rangle$, nearest-neighbor density-density correlation $\langle n_i n_j \rangle$ for $ \mbox{\boldmath $r$}_{ij}$ at $\pm$30 -degree angles with respect to horizontal axis (i.e., parallel to $t_1$, $t_2$, and $t_3$ bonds), and $\langle n_i n_j \rangle$ for $ \mbox{\boldmath $r$}_{ij}$ parallel to vertical axis (i.e., parallel to $t_4$ bonds), as functions of $V_2$. (c) Optical conductivity spectra for $V_2=0.35$ with different polarizations as indicated. The peak-broadening parameter is set at 0.005. 
\label{fig:bnt_btr16_trat1t08t02t01u8v3vxx_eq}}
\end{figure}

Now we consider the triangular lattice with bent trimers shown in Fig.~\ref{fig:latt_str}(b). To see how the intersite repulsive interactions $V_1$ and $V_2$ govern the ground states, we show the electron density distribution $\langle n_i \rangle$ in Fig.~\ref{fig:bnt_btr16_trat1t08t02t01u8v3vxx_eq}(a) and the spatially averaged correlation functions $\langle n_{i\uparrow} n_{i\downarrow} \rangle$ and $\langle n_i n_j \rangle$ in Fig.~\ref{fig:bnt_btr16_trat1t08t02t01u8v3vxx_eq}(b) as functions of $V_2$. 
Sites A and A' are again crystallographically equivalent [although the symmetry is different from that in Fig.~\ref{fig:latt_str}(a)], so that their electron densities are equal in the ground states. Most of the behaviors are similar to those in the previous section once the bonds are classified as vertical and nonvertical. Figure~\ref{fig:bnt_btr16_trat1t08t02t01u8v3vxx_eq}(a) is similar to Fig.~\ref{fig:lnr_atr16_trat1413020603u8v3vxx_eq}(a). In Fig.~\ref{fig:bnt_btr16_trat1t08t02t01u8v3vxx_eq}(b), $\langle n_i n_j \rangle$ for $ \mbox{\boldmath $r$}_{ij}$ at a 30 -degree angle with respect to the horizontal axis and that at a $-$30 -degree angle are identical because of the symmetry  with respect to this axis. Otherwise, Fig.~\ref{fig:bnt_btr16_trat1t08t02t01u8v3vxx_eq}(b) is similar to Fig.~\ref{fig:lnr_atr16_trat1413020603u8v3vxx_eq}(b). Since we mainly use $V_2=0.35$ for photoinduced dynamics below, we show the optical conductivity spectra for $V_2=0.35$ in Fig.~\ref{fig:bnt_btr16_trat1t08t02t01u8v3vxx_eq}(c). The main charge-transfer excitations are seen below 0.4, so that $ \omega=0.8 $ for photoexcitation used below is well above them. 

\subsection{Densities and correlations after pulse excitation \label{sec:bent}}
\begin{figure}
\includegraphics[height=16.8cm]{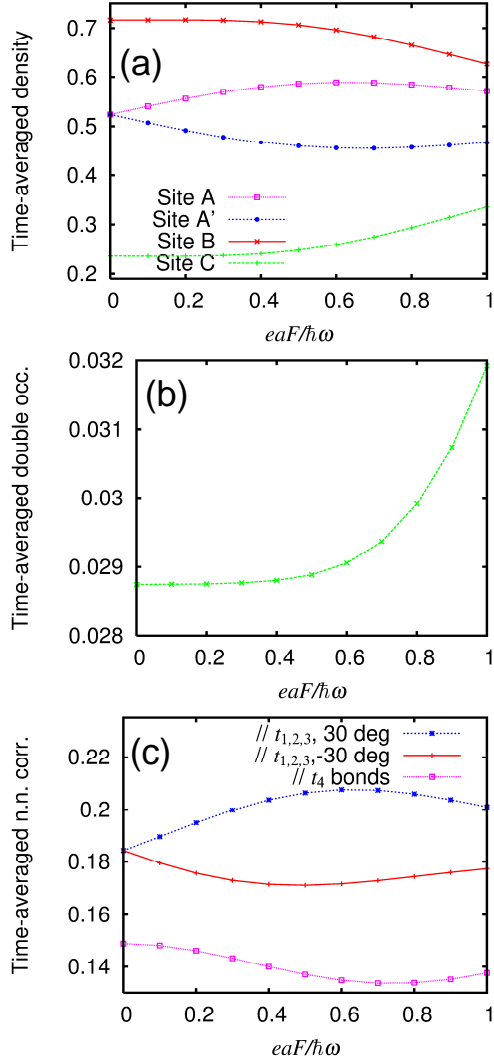}
\caption{(Color online) 
(a) Time-averaged electron densities $\langle \langle n_i \rangle \rangle$ at sites A, A', B, and C, (b) spatially and temporally averaged double occupancy $\langle \langle n_{i\uparrow} n_{i\downarrow} \rangle \rangle$, and (c) spatially and temporally averaged nearest-neighbor density-density correlations $\langle \langle n_i n_j \rangle \rangle$ for $ \mbox{\boldmath $r$}_{ij}$ in different directions as indicated, for $V_2=0.35$ and $\theta=-90\deg$, as functions of $ eaF/(\hbar\omega) $. 
\label{fig:bnt_btr16_trat1t08t02t01u8v3v35_w0p80n1fxqm90}}
\end{figure}

Unless stated otherwise, we use $\theta=-90\deg$ in Sect.~\ref{sec:bent}, which corresponds to the direction parallel to the line connecting the end points of a bent trimer A-B-A' (and also of a weak trimer A-C-A'). The time-averaged electron density distribution $\langle \langle n_i \rangle \rangle$ is shown in Fig.~\ref{fig:bnt_btr16_trat1t08t02t01u8v3v35_w0p80n1fxqm90}(a). 
The behavior in the range of $ 0 < eaF/(\hbar\omega) < 1 $ in Fig.~\ref{fig:bnt_btr16_trat1t08t02t01u8v3v35_w0p80n1fxqm90}(a) is roughly similar to that in the range of $ 0 < eaF/(\hbar\omega) < 0.5 $ in Fig.~\ref{fig:lnr_atr16_trat1413020603u8v3v35_w0p80n1fxq00}(a). The equivalence of sites A and A' is broken by the field. In the present case, $\langle \langle n_i \rangle \rangle$ at site A and $\langle \langle n_i \rangle \rangle$ at site A' are exchanged for $\theta \rightarrow -\theta$. The averaged double occupancy $\langle \langle n_{i\uparrow} n_{i\downarrow} \rangle \rangle$ shown in Fig.~\ref{fig:bnt_btr16_trat1t08t02t01u8v3v35_w0p80n1fxqm90}(b) behaves differently from that in Fig.~\ref{fig:lnr_atr16_trat1413020603u8v3v35_w0p80n1fxq00}(b): it monotonically increases with the field amplitude. For small field amplitudes, however, the change rate is quite small. There exist polarizations for which $\langle \langle n_{i\uparrow} n_{i\downarrow} \rangle \rangle$ decreases for small field amplitudes, but the change rate in that case is also quite small (not shown). Note that the range of the ordinate in Fig.~\ref{fig:bnt_btr16_trat1t08t02t01u8v3v35_w0p80n1fxqm90}(b) is  one order of magnitude narrower than that in Fig.~\ref{fig:lnr_atr16_trat1413020603u8v3v35_w0p80n1fxq00}(b). 

The behavior of the averaged nearest-neighbor density-density correlations $\langle \langle n_i n_j \rangle \rangle$ shown in Fig.~\ref{fig:bnt_btr16_trat1t08t02t01u8v3v35_w0p80n1fxqm90}(c) is similar to that in Fig.~\ref{fig:lnr_atr16_trat1413020603u8v3v35_w0p80n1fxq00}(c). Because of $V_2 > V_1$, $\langle \langle n_i n_j \rangle \rangle$ along the $t_4$ bonds is smaller than the others already before photoexcitation (at $F=0$), and it is further decreased by electric-field pulses. The anisotropy in $\langle \langle n_i n_j \rangle \rangle$ is enhanced by them, which is interpreted as the photoinduced enhancement of the anisotropy in the {\it effective} intersite repulsive interactions, even for the present structure with bent trimers. 

\begin{figure}
\includegraphics[height=5.6cm]{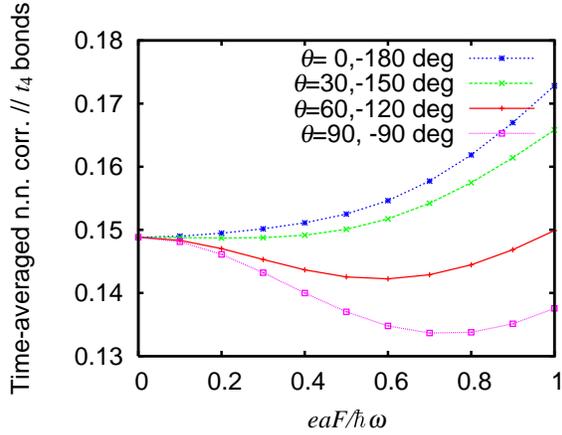}
\caption{(Color online) 
Spatially and temporally averaged nearest-neighbor density-density correlation $\langle \langle n_i n_j \rangle \rangle$ for $ \mbox{\boldmath $r$}_{ij}$ parallel to vertical axis (i.e., parallel to $t_4$ bonds), for $V_2=0.35$ and different polarizations of photoexcitation $\theta$, as functions of $ eaF/(\hbar\omega) $. Here, the data for $\theta$ and $\theta+180\deg$ are averaged. 
\label{fig:bnt_btr16_trat1t08t02t01u8v3v35_w0p80n1fxqyy_avnn}}
\end{figure}

The polarization ($\theta$) dependence of $\langle \langle n_i n_j \rangle \rangle$ along the $t_4$ bonds is shown in Fig.~\ref{fig:bnt_btr16_trat1t08t02t01u8v3v35_w0p80n1fxqyy_avnn}. 
In the present structure with bent trimers, the inversion symmetry does not exist, in contrast to the previous structure with linear trimers possessing this symmetry. Then, the data for $\theta$ and $\theta+180\deg$, which would be difficult to distinguish experimentally, are averaged and made easy to see. At $\theta = \pm90\deg$ (the data for $\theta = 90\deg$ and $\theta = -90\deg$ are actually identical), $\langle \langle n_i n_j \rangle \rangle$ becomes smallest for any $ eaF/(\hbar\omega) $. This polarization is parallel to the line connecting the end points of two types of trimers (linked by $t_1$ and by $t_2$). In this sense, the present result is similar to the previous one for linear trimers: the $t_2$ process efficiently uses site C where the electron density is lowest in the ground state. For $\theta = 60\deg$, $\langle \langle n_i n_j \rangle \rangle$ is also decreased by the field in the entire range displayed. For $\theta = 30\deg$, $\langle \langle n_i n_j \rangle \rangle$ slightly decreases for very small field amplitudes; it increases otherwise. For $\theta = 0\deg$,  $\langle \langle n_i n_j \rangle \rangle$ monotonically increases with the field amplitude. Note that photoexcitation with $\theta = 0\deg$ does not break the reflection symmetry, and that sites A and A' remain crystallographically equivalent for $\theta = 0\deg$. This implies that the photoinduced enhancement of the anisotropy requires an escape from the crystallographic equivalence. If directionally averaged over $\theta$, $\langle \langle n_i n_j \rangle \rangle$ is decreased by not-too-strong fields. 

\begin{figure}
\includegraphics[height=11.2cm]{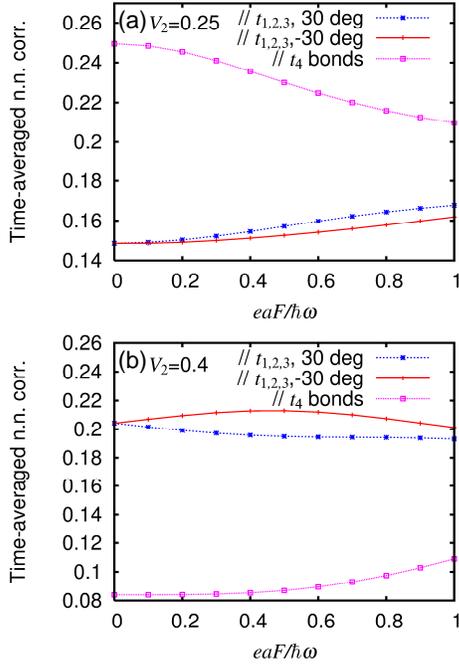}
\caption{(Color online) 
Spatially and temporally averaged nearest-neighbor density-density correlations $\langle \langle n_i n_j \rangle \rangle$ for $ \mbox{\boldmath $r$}_{ij}$ in different directions as indicated, for $\theta=-90\deg$, (a) $V_2=0.25$ and (b) $V_2=0.4$, as functions of $ eaF/(\hbar\omega) $. 
\label{fig:bnt_btr16_trat1t08t02t01u8v3v25v4_w0p80n1fxqm90_avnn3}}
\end{figure}

Finally, $\langle \langle n_i n_j \rangle \rangle$ for $V_2=0.25$ and $V_2=0.4$ are shown in Figs.~\ref{fig:bnt_btr16_trat1t08t02t01u8v3v25v4_w0p80n1fxqm90_avnn3}(a) and \ref{fig:bnt_btr16_trat1t08t02t01u8v3v25v4_w0p80n1fxqm90_avnn3}(b), respectively. 
In both cases, $\langle \langle n_i n_j \rangle \rangle$ that is larger before photoexcitation basically decreases, and $\langle \langle n_i n_j \rangle \rangle$ that is smaller before photoexcitation increases, as the field amplitude increases. Thus, the anisotropy in $\langle \langle n_i n_j \rangle \rangle$ is weakened by photoexcitation, in a manner similar to the previous one for linear trimers. These behaviors are in contrast to the photoinduced enhancement of the anisotropy in  $\langle \langle n_i n_j \rangle \rangle$ for $V_2=0.35$ shown in Fig.~\ref{fig:bnt_btr16_trat1t08t02t01u8v3v35_w0p80n1fxqm90}(c). In any case, a fact that is more important is the photoinduced enhancement of anisotropic charge correlations, which is caused by $V_2$ being slightly larger than $V_1$ on different lattices with trimers. 

\section{Conclusions and Discussion}
To explore the feasibility of photoinduced charge localization and nontrivial modulations of charge correlations, we treat quarter-filled extended Hubbard models on triangular lattices with trimers, where the end points are crystallographically equivalent. To clarify the generality of the findings, we employ a triangular lattice with linear trimers like that of $\alpha$-(BEDT-TTF)$_2$I$_3$ in the metallic phase and another lattice with bent trimers. For different relative strengths of anisotropic intersite interactions, we calculate photoinduced dynamics using the exact diagonalization method and the time-dependent Schr\"odinger equation. 

For the triangular lattice with linear trimers, the spatially and temporally averaged double occupancy decreases after not-too-strong electric-field pulses (even for other polarizations $\theta$ and for $V_2 = 0.25$ and $V_2 = 0.4$ in addition to the data presented above, not shown) are applied as if the on-site repulsion $U$ were increased relative to the transfer integrals transiently after photoexcitation. In $\alpha$-(BEDT-TTF)$_2$I$_3$, the intersite repulsion $V_2$ is known to be slightly larger than $V_1$ owing to the difference in the intermolecular distance. In this case, the spatially and temporally averaged nearest-neighbor density-density correlations behave as if the intersite repulsion $V_2$ were increased and $V_1$ were decreased relative to the transfer integrals after photoexcitation. 

In the transient state experimentally observed immediately after photoexcitation,\cite{ishikawa_ncomms14} the lattice is not yet expected to be sufficiently distorted to stabilize long-range charge order, and the optical freezing of charge motion lasts for only 50 fs, so that the order, if any, would be short-ranged. In this sense, the numerical calculations performed on finite-size systems for ultrafast dynamics without coupling to a heat bath would be suitable. The theoretically found, enhanced {\it effective} interactions $U$ and $V_2$ relative to the transfer integrals are consistent with the experimentally observed optical freezing of charge motion.\cite{ishikawa_ncomms14} 

The lattice structures with trimers studied in this paper allow the presence of a charge-rich site B at the center and the crystallographically equivalent end points A and A' of a trimer. The intersite repulsion $V_2$ is substantial between these sites [A and A' belong to neighboring trimers in Fig.~\ref{fig:latt_str}(a) and to a single trimer in Fig.~\ref{fig:latt_str}(b)]. In both cases of linear and bent trimers, the photoinduced enhancement of anisotropic charge correlations is observed only when $V_2$ is slightly larger than $V_1$. 

The condition on $V_1$ and $V_2$ is understood from the following points. (i) Before photoexcitation, the end points of a trimer are equivalent, so that their electron densities are equal. (ii) The intersite repulsion $V_2$ is substantial between them, so that they are ready to disproportionate. Charge order may emerge as a consequence of freedom from oppression, i.e., crystallographic equivalence. (iii) The emergence is triggered by a rather small perturbation such as coupling to lattice degrees of freedom (as realized in equilibrium below the charge-order metal-insulator transition temperature,\cite{tanaka_jpsj08,miyashita_jpsj08,yamamoto_jpsj08} which is an example of electronic ferroelectricity\cite{ishihara_jpsj10}) or an electric field. Ultrafast photoexcitation does not allow the lattice to be substantially distorted, but it can prevent thermalization at an early stage. If $V_2$ is small, the condition (ii) is not fulfilled. If $V_2$ is too large (but substantially smaller than $U$), the strong density-density correlation between sites A and A', which is already developed before photoexcitation, is merely weakened by photoexcitation. Therefore, it is an issue of balance. A trimer works indeed as a balancing toy that accepts largely inclined states. To coherently incline balancing toys, we need interactions between them and external stimuli. 

From the above discussions, the existence of crystallographically equivalent sites is found to be essential. Indeed, even if we distort the triangular lattices to make square lattices with diagonal lines in one direction, keeping the crystallographic equivalence of sites A and A', we find very similar phenomena (not shown). Therefore, this is different from a charge-ordered liquid state for spinless fermions on triangular lattices,\cite{hotta_prb06,hotta_jpsj06} which has been discussed in the context of geometrical frustration in triangular lattices as realized in $\theta$-(BEDT-TTF)$_2X$ compounds in the metallic phase. 

From these discussions, the optical freezing of charge motion observed in $\alpha$-(BEDT-TTF)$_2$I$_3$\cite{ishikawa_ncomms14} is presumably attributed to the competing intersite repulsive interactions and the lattice structure with trimers possessing crystallographically equivalent end points. It is not as simple as dynamical localization. Although the importance of interactions has been suggested,\cite{ishikawa_ncomms14,naitoh_prb16} a mechanism is made clearer after considering lattice structures and the competition of interactions. For $V_2=0.35$ and linear trimers, the decrease in the averaged double occupancy for $\theta=0$ and $eaF/(\hbar\omega)=0.4$ is reproduced in the ground state by the increase in $U$ relative to the bandwidth by 4\%, while the theory for dynamical localization leads to the increase in $U$ relative to the renormalized bandwidth by 3\% at the same parameters for $\theta$ and $eaF/(\hbar\omega)$. However, the photoinduced enhancement of anisotropic charge correlations is not explained by the theory for dynamical localization or an extension of it within the framework of the quantum Floquet theory, which will be reported elsewhere. A similar behavior regarding the photoinduced enhancement of the anisotropy is realized even if trimers are bent as long as the end points are crystallographically equivalent. It would be experimentally difficult to verify the mechanism, but it would be useful to see the dependence of the phenomenon on the polarization of photoexcitation. For the present triangular lattices consisting of equilateral triangles, the photoinduced enhancement of anisotropic charge correlations is almost maximized when polarized parallel to the line connecting the end points of a weak trimer. For specific materials, we would need a detailed study of the polarization dependence. 

\begin{acknowledgment}
The author is grateful to S. Iwai for continual discussions from the discovery of the phenomenon and for Y. Tanaka for various theoretical discussions. 
This work was supported by Grants-in-Aid for Scientific Research (C) (Grant No. 16K05459) and Scientific Research (A) (Grant No. 15H02100) from the Ministry of Education, Culture, Sports, Science and Technology of Japan. 
\end{acknowledgment}

\bibliography{67911}

\end{document}